\documentclass[twocolumn,showpacs,preprintnumbers,amsmath,amssymb,superscriptaddress]{revtex4}
\usepackage[bookmarks=false]{hyperref}
\usepackage{amssymb}
\usepackage{amsmath}
\usepackage{graphicx}
\usepackage{dcolumn}
\usepackage{bm}
\usepackage{epstopdf}
\usepackage{times}
\usepackage{color}

\newcommand{\PFA}{PrFe$_2$Al$_8$}
\begin{document}
\preprint{}
%
\title{Pr-magnetism in the quasi-skutterudite compound PrFe$_2$Al$_8$}
%
%
%
\author{Harikrishnan S. Nair}
\email{h.nair.kris@gmail.com, hsnair@colostate.edu}
\affiliation{Highly Correlated Matter Research Group, Physics Department, P. O. Box 524, University of Johannesburg, Auckland Park 2006, South Africa}
\affiliation{Department of Physics, Colorado State University, 200 W. Lake St., Fort Collins, CO 80523-1875, USA}
\author{Michael O. Ogunbunmi}
\affiliation{Highly Correlated Matter Research Group, Physics Department, P. O. Box 524, University of Johannesburg, Auckland Park 2006, South Africa}
\author{C. M. N. Kumar}
\affiliation{Laboratoire National des Champs Magn\'{e}tiques Intenses, CNRS-INSA-UJF-UPS, 143 Avenue de Rangueil, F-31400 Toulouse, France}
\author{D. T. Adroja}
\affiliation{ISIS Facility, STFC, Rutherford Appleton Laboratory, Chilton, Didcot, Oxfordshire OX11 0QX, United Kingdom}
\author{P. Manuel}
\affiliation{ISIS Facility, STFC, Rutherford Appleton Laboratory, Chilton, Didcot, Oxfordshire OX11 0QX, United Kingdom}
\author{D. Fortes}
\affiliation{ISIS Facility, STFC, Rutherford Appleton Laboratory, Chilton, Didcot, Oxfordshire OX11 0QX, United Kingdom}
\author{J. Taylor}
\affiliation{ISIS Facility, STFC, Rutherford Appleton Laboratory, Chilton, Didcot, Oxfordshire OX11 0QX, United Kingdom}
\author{Andr\'{e} M. Strydom}
\affiliation{Highly Correlated Matter Research Group, Physics Department, P. O. Box 524, University of Johannesburg, Auckland Park 2006, South Africa}
\begin{abstract}
The intermetallic compound \PFA\ that possesses 
a three-dimensional network structure of Al polyhedra 
centered at the transition metal element Fe and 
the rare earth Pr is investigated through neutron 
powder diffraction and inelastic neutron scattering 
in order to elucidate the magnetic ground state of Pr and Fe
and the crystal field effects of Pr. Our neutron diffraction study 
confirms long-range magnetic order of Pr below $T_N $ = 4.5~K
in this compound. Subsequent magnetic structure 
estimation reveals a magnetic propagation vector 
$\bf k$ = {\bf ($\frac{1}{2}$~0~$\frac{1}{2}$)} with a magnetic 
moment value of 2.5~$\mu_\mathrm{B}$/Pr along 
the orthorhombic $c$-axis and evidence the lack of 
ordering in the Fe sublattice. The inelastic neutron 
scattering study reveals one crystalline electric field 
excitation near 19~meV at 5~K in \PFA. The energy-integrated 
intensity of the 19~meV excitation as a function of $|Q| (\AA^{-1})$ 
follows the square of the magnetic form factor of Pr$^{3+}$ 
thereby confirming that the inelastic excitation belongs to the 
Pr sublattice. The second sum rule applied to the dynamic structure 
factor indicates only 1.6(2)~$\mu_\mathrm{B}$ evolving at 
the 19~meV peak compared to the 3.58~$\mu_\mathrm{B}$ 
for free Pr$^{3+}$, indicating that the crystal field  ground state is 
magnetic and the missing moment is associated with the 
resolution limited quasi-elastic line. The magnetic order 
occurring in Pr in \PFA\ is counter-intuitive to the 
symmetry-allowed crystal field level scheme, hence, is 
suggestive of exchange-mediated mechanisms of ordering 
stemming from the magnetic ground state of the crystal field levels.
\end{abstract}
\maketitle
\section{\label{INTRO}Introduction}
Intermetallic compounds possessing a cage-like crystal 
structure comprising of polyhedra that contain rare 
earth or transition metal atoms in their voids have 
been interesting compounds from the perspective 
of thermoelectric properties, however, the magnetism 
of rare earth that is placed in the cages is a relatively 
less-explored topic. The filled-skutterudites $RT_4X_{12}$ 
where $R$ = rare earth, $T$ = transition metal and 
$X$ = pnictogen, is a well-explored class of compounds 
\cite{sales_filled_1996,nolas_skutterudites:_1999} 
with the potential of making thermoelectric applications
\cite{guo_development_2012}. The original binary skutterudite 
formula can be written as $TX_3$ which, with the existence 
of voids in the original $Im3$ structure, may adopt a general 
formula $R_2M_8X_{24}$ or $RM_4X_{12}$. The cubic 
$RT_2$Al$_{20}$, known as the Frank Kasper compounds, 
provide another example of the caged-structure compounds 
similar to the filled-skutterudites, and have a large 
coordination number of atoms constructing the cage 
around the $R$ atom\cite{frank_complex_1958}. They 
crystallize in the space group $Fd \overline{3}m$ 
($\#$227) with the $R$ ion surrounded by neighboring 
16 atoms and the $R$ site is of a cubic point 
symmetry. A related caged-structure compound is the 
$RT_2$Al$_{10}$ which has orthorhombic crystal 
structure with the space group $Cmcm$ ($\#$63)
\cite{tursina_ceru2al10_2005,adroja_muon-spin-relaxation_2013}. 
In this caged system the $R$ atoms are 20-fold coordinated 
and are located inside $T-$Al cages. It is interesting to 
look at the magnetism of these caged-network compounds 
with Pr occupying the void inside the cage. For example, 
the filled-skutterudite PrRu$_4$Sb$_{12}$ shows a 
Van Vleck paramagnetic behaviour for most of the 
temperature range before entering a superconducting 
phase below 1.04~K\cite{takeda_superconducting_2000,adroja_probing_2005}. 
In the case of PrTi$_2$Al$_{20}$, a non-magnetic order 
parameter was identified below the magnetic transition at 
2~K\cite{ito_sr_2011}. This was correlated with the 
ferroquadrupole ordering scenario. In the case of 
PrRu$_2$Al$_{10}$, which displayed metallic behaviour, 
no magnetic phase transition was recorded down 
to low temperature\cite{sakoda_single_2012}. 
\begin{figure}[!t]
\centering
\includegraphics[scale=0.075]{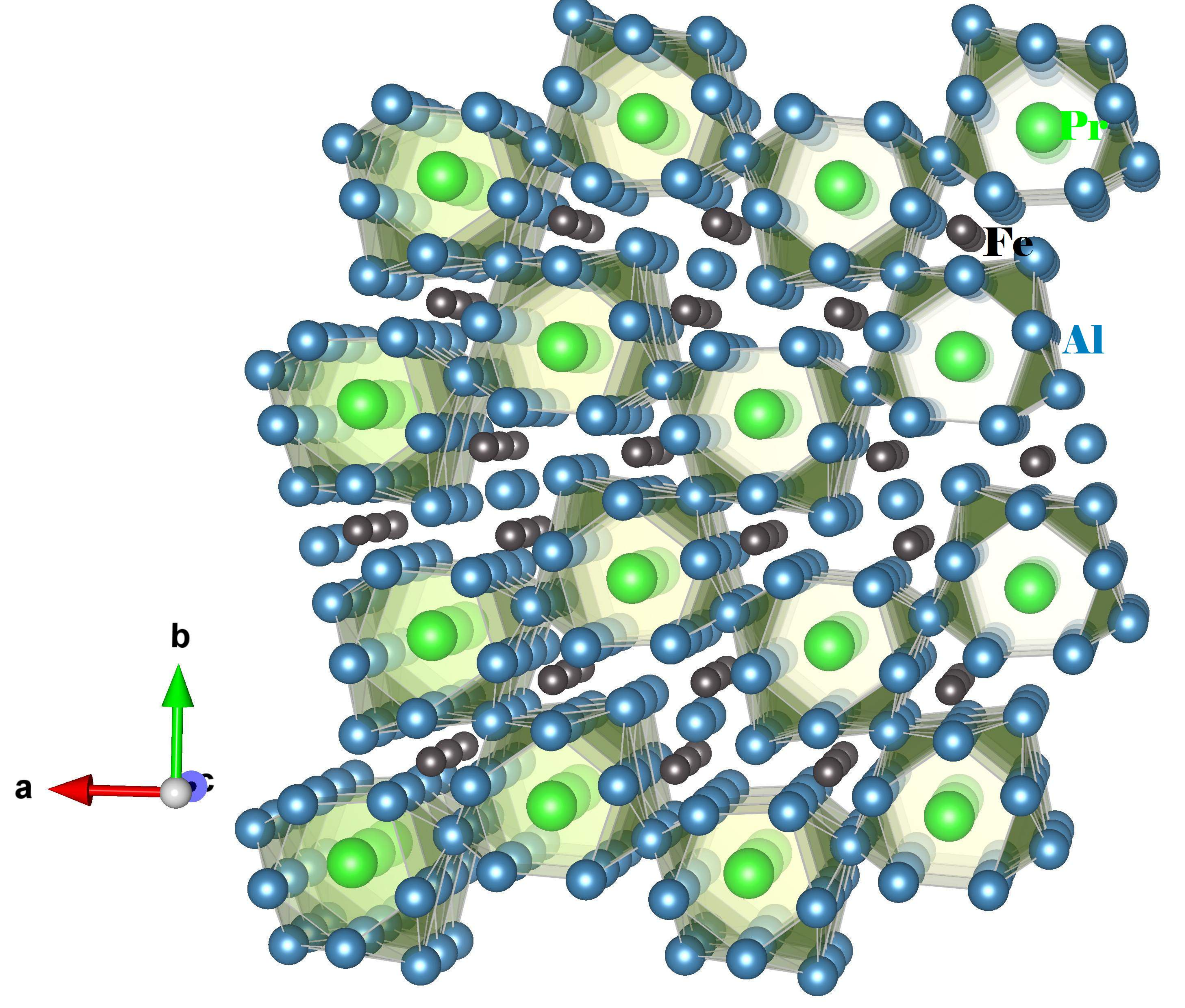}
\caption{(color online) A schematic representation of the cage-like 
network structure of the quasi-skutterudite \PFA. A projection of the 
cages on the $ab$ plane is shown here. Note that the representation is 
not a unit cell. The green spheres represent Pr while the dark gray are 
Fe and the turquoise are Al. \label{fig_str}}
\end{figure}
These studies focusing on the rare earth magnetism in 
the caged-network points towards the interesting 
magnetic ordering or the lack of it, that may have a 
strong influence from the ligand field around the rare earth. 
The crystal field ground state of the $4f$ ion in the cages
play an important role in the magnetic ordering phenomena
in these compounds.
While many studies focused on the magnetism of the $4f$ element
in the cages, it must be noted that the magnetism of the $3d$
element and the correlation between the two is an emergent theme
in similar compounds. For example, {\em ab-initio} calculations
in the Ce$_3T_4X_{13}$ compounds have shown indication of
correlation between the $3d$ and the $4f$ wave functions\cite{slebarski2015study}.
Unexpected localization of Fe $3d$ moments were experimentally observed
in another caged compound Yb$_2$Fe$_2$Al$_{10}$ where it
was observed that the long-range ordering of the $3d$ moments
were avoided down to 0.4~K\cite{khuntia2014contiguous}.
These studies point to case of caged intermetallics where
the rare earth and transition metal display cooperative 
phenomena in concert or as separate.
With this background, we take a look at a rather 
less-studied caged-network intermetallic compound, 
PrFe$_2$Al$_8$ that belongs to the family of $RT_2X_8$ ($X$ = Al, Ga).
\\
\begin{table}[!b]
\centering 
\caption{The lattice parameters of \PFA\ obtained from the refinement of 
	neutron diffraction data at 300, 150 and 25~K are provided in the top 
	section of the table. The refined atomic coordinates of \PFA\ at 300~K 
	are given in the bottom section. These values are obtained from the 
	HRPD-data. \label{tab1}}
\setlength{\tabcolsep}{10pt}
\begin{tabular}{llll} \hline\hline
&        300~K          &      150~K        &        25~K     \\ \hline\hline
$a$ ({\AA})          &        12.5101(3)     &     12.4871(5)     &      12.4799(4)   \\  
$b$ ({\AA})          &       14.4383(3)      &     14.4168(6)     &      14.4081(8)     \\
$c$ ({\AA})          &       4.0411(1)       &     4.0332(8)      &      4.0297(2)   \\ \hline
\end{tabular} \\
\PFA, $Pbam$, 300~K\\
\begin{tabular}{lccccc}\hline
Atom & site & $x$  & $y$  &    $z$   & $U_{iso}$ \\ \hline\hline  
Pr   & $4g$  & 0.34430 &  0.31793 & 0.0 & 0.0029 \\
Fe(1)   & $4g$  & 0.15095 &  0.09235 & 0.0 & 0.0018 \\
Fe(2)   & $4g$  & 0.03619 & 0.40434 & 0.0 & 0.0021 \\
Al(1)   & $2d$  & 0.0000  & 0.5000 & 0.5 & 0.0033 \\
Al(2)   & $2a$  & 0.0000  & 0.0000 & 0.0 & 0.0053 \\
Al(3)   & $4h$  & 0.02968 &  0.13355 & 0.5 & 0.0053 \\
Al(4)   & $4h$  & 0.16160 &  0.3761 & 0.5 & 0.0031 \\
Al(5)   & $4h$  & 0.23677 &  0.17102 & 0.5 & 0.0035 \\
Al(6)   & $4h$  & 0.33260 &  0.49029 & 0.5 & 0.0034 \\
Al(7)   & $4h$  & 0.45987 &  0.18346 & 0.5 & 0.0028 \\
Al(8)   & $4g$  & 0.33473 &  0.03719  & 0.0 & 0.0041 \\
Al(9)   & $4g$  & 0.09153 &  0.25152  & 0.0 & 0.0043 \\ \hline\hline
\end{tabular} 
\end{table}
\indent
The $RT_2X_8$  compounds crystallize in cubic 
$Pbam$ space group. Figure~\ref{fig_str} shows the 
projection of the crystal structure on $ab$-plane displaying 
the network of Al and the chain-like formation of Pr and Fe 
along the $c$-axis. These structural peculiarities certainly 
have a bearing on the magnetism of this class of compounds, 
notably, a strong influence from crystal field effects. 
Depending on whether the crystal field levels are 
comparable in energy scales with other interactions 
(especially, magnetic exchange), a non-magnetic singlet 
or doublet crystal field ground state can result. To cite 
an example, the point charge model calculations that take 
in to account the local symmetry for the Pr site in PrSi 
predicts a splitting of $J$ = 4 Pr$^{3+}$ ground state in to 
9 singlets which renders spontaneous magnetic order impossible. 
However, an anomalous ferromagnetic ground state 
(with an ordering temperature of $\approx$ 55~K) was 
experimentally unraveled in PrSi\cite{snyman2012anomalous}. 
Similarly, the point symmetry of the Pr-site in \PFA\ is 
$m$ ($C_s$) which predicts 9 distinct singlets from a 9-fold 
degenerate state which consequentially presupposes a 
non-magnetic ground state. Our previous studies on 
\PFA\ had revealed long-range magnetic order through 
macroscopic magnetization measurements which was 
assumed to have originated from the Pr magnetic 
sublattice\cite{nair2016magnetic}. The Fe sublattice was 
deduced to be non-magnetic or as having only short-range 
spin correlations. In the present work, we investigate the 
magnetic ground state of \PFA\ through neutron diffraction 
experiments and confirm that only Pr orders magnetically long-range
while Fe is not. In addition, our study sheds light on the magnetic excitations 
and crystal field effects through inelastic neutron scattering.
\begin{table}[!t]
\centering 
\caption{The Pr-Pr and the Fe-Fe nearest-neighbour distances and angles of \PFA\ at 300~K, 150~K, 25~K and 1.6~K. The distances along $a$, $b$ and $c$ directions are shown. The angle shown is that which is formed between any three nearest-neighbour Pr atoms.\label{tab2}}
\setlength{\tabcolsep}{3pt}
\begin{tabular}{lccccc}\hline
& 300~K             & 150~K            &   25~K            &   1.6~K          \\ \hline\hline  
Pr-Pr ($\parallel c$)   & 4.0411(9) {\AA}  & 4.0334(6) {\AA}  &  4.0297(2) {\AA} &  4.0297(2) {\AA} \\
Pr-Pr ($\parallel a$)   & 6.572(9) {\AA}   & 6.562(11) {\AA}  &  6.559(11) {\AA} &  6.554(13) {\AA} \\
Pr-Pr ($\parallel b$)   & 7.563(7) {\AA}   & 7.554(7) {\AA}   &  7.551(9) {\AA}  &  7.560(9) {\AA} \\ \hline
Fe-Fe ($\parallel c$)   & 4.0411(2) {\AA}  & 4.0335(6) {\AA}  &  4.0297(6) {\AA} &  4.0297(8) {\AA} \\
Fe-Fe ($\parallel a$)   & 4.818(14) {\AA}  & 4.815(4) {\AA}  &  4.816(5) {\AA} &  4.817(6) {\AA} \\
Fe-Fe ($\parallel b$)   & 2.754(4) {\AA}  & 2.748(5) {\AA}  &  2.757(5) {\AA} &  2.762(7) {\AA} \\ \hline
$\angle$ Pr ($\parallel a$)   & 144.23(8)$^\circ$  & 144.12(7)$^\circ$  & 144.12(7)$^\circ$ &  144.36(9)$^\circ$ \\
$\angle$ Pr ($\parallel b$)   & 145.32(9)$^\circ$  & 145.19(11)$^\circ$  & 145.15(10)$^\circ$ &  145.68(11)$^\circ$ \\ \hline
\end{tabular} 
\end{table}
\section{\label{EXP}Experimental details}
About 10 g of polycrystalline powder of 
\PFA\ and LaFe$_2$Al$_8$ were prepared 
following the procedure outlined in Ref.\cite{nair2016magnetic}. 
The powder samples were initially characterized using 
laboratory x-rays to check for purity of the chemical 
phase and using magnetometry to identify the magnetic 
phase transition. These preliminary checks confirmed 
the findings of Ref.\cite{nair2016magnetic}.
Time-of-flight (TOF) neutron powder diffraction 
experiments were carried out at the instrument 
HRPD at ISIS facility of the 
Rutherford Appleton Laboratory, UK mainly to characterize
the temperature evolution of the crystal structure. 
Diffraction patterns covering a $d$-spacing range 
0.75 - 4~{\AA} were obtained at 19 temperature 
points in the interval of 300~K to 1.6~K. Additionally, 
experiments were also performed using the TOF 
diffractometer WISH on TS2 at ISIS facility 
especially covering the higher $d$-spacing range of 
1.2 - 11~{\AA} in order to unambiguously determine 
the magnetic structure of \PFA.
Inelastic neutron scattering experiments were 
performed on 10~g of \PFA\ powder sample using 
the TOF chopper spectrometer MARI at ISIS facility. 
\PFA~ powder from the same batch was used in all 
the three experiments. Incident neutron energies 
$E_i$ = 8~meV, 50~meV and 150~meV were selected 
for the experiment. The chopper frequencies were set at 
150~Hz, 250~Hz,  and 400~Hz, respectively. Data were 
collected mainly at 5~K, 50~K, 200~K and at 300~K. 
As a non-magnetic reference, the inelastic response of 
LaFe$_2$Al$_8$ was also recorded to serve as a 
phonon estimate for \PFA. The inelastic neutron 
data was analyzed using the software DAVE\cite{dave}. 
Rietveld refinement\cite{rietveld} of the diffraction 
data were carried out using FULLPROF software
\cite{fullprof}. The software SARA$h$\cite{sarah} 
was used for the representation analysis of magnetic 
structure from diffraction data.
\section{\label{RESULTS} Results}
\subsection{Neutron Diffraction}
\begin{figure}[!t]
\centering
\includegraphics[scale=0.28]{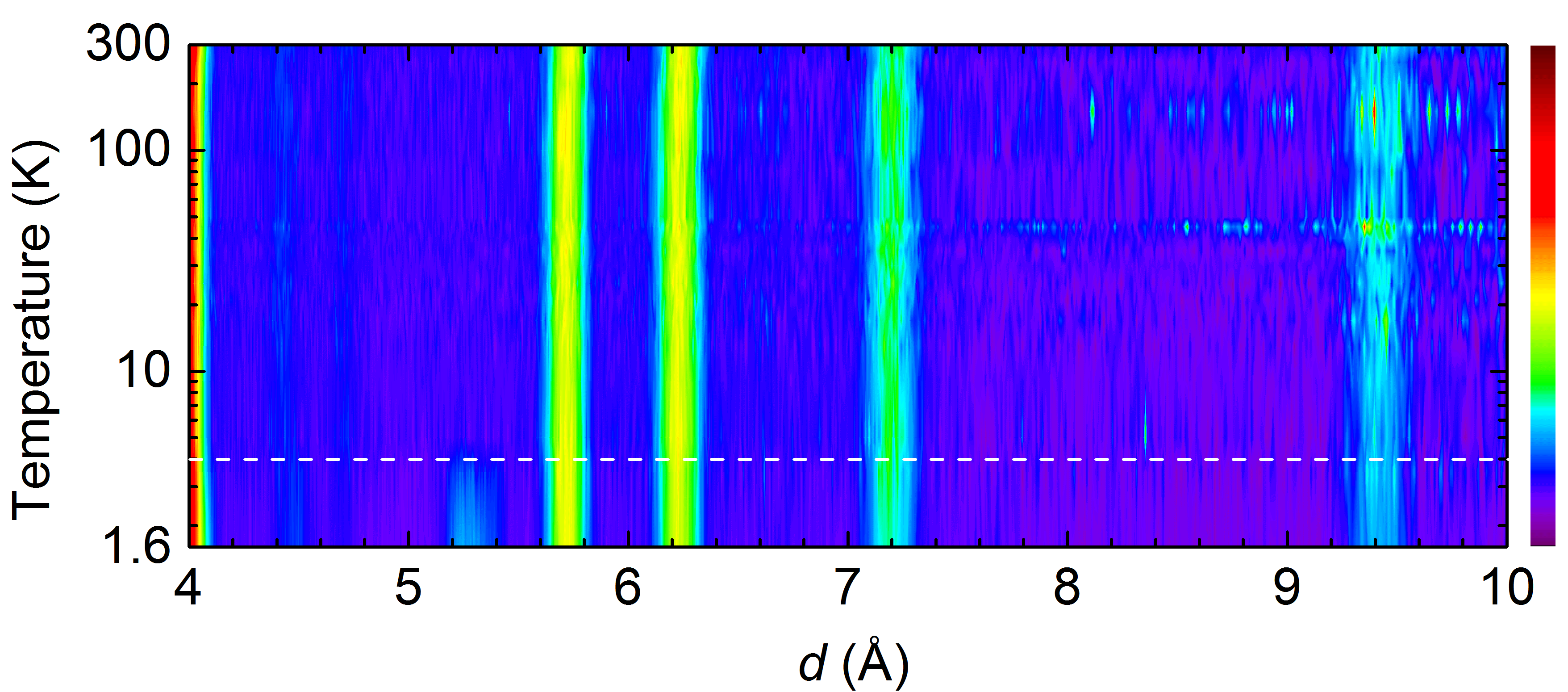}
\caption{(color online) A colour plot of diffracted intensity from 
\PFA\, measured at HRPD, plotted as a function of temperature 
versus $d$-spacing. The temperature axis is shown in log scale. 
A horizontal dashed line marks $T^\mathrm{Pr}_N \approx$ 4~K where 
the magnetic long-range order of Pr begins as evidenced by the 
development of intensity at $d \approx$ 5.3~{\AA}. \label{fig_hrpd_contour}}
\end{figure}

\begin{figure}[!t]
\centering
\includegraphics[scale=0.34]{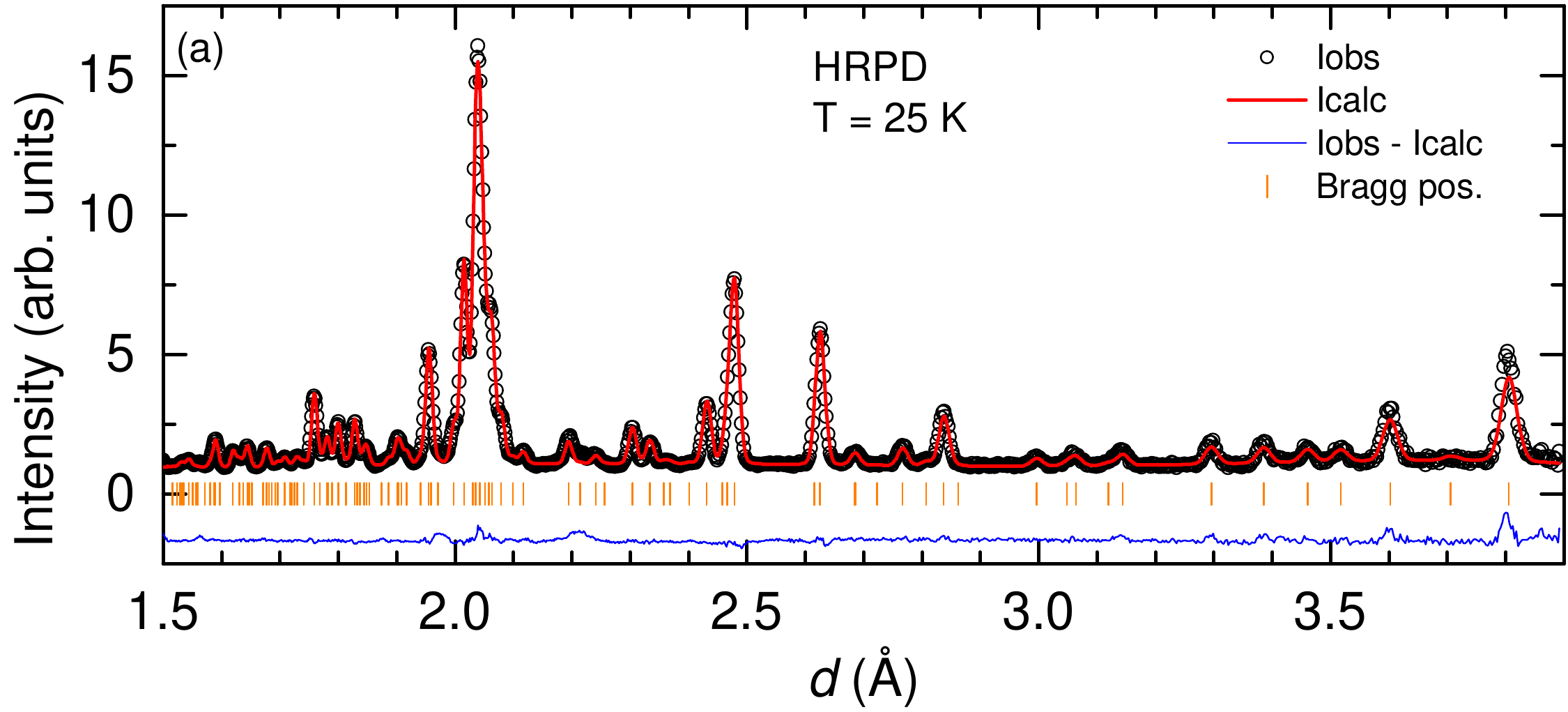}
\includegraphics[scale=0.35]{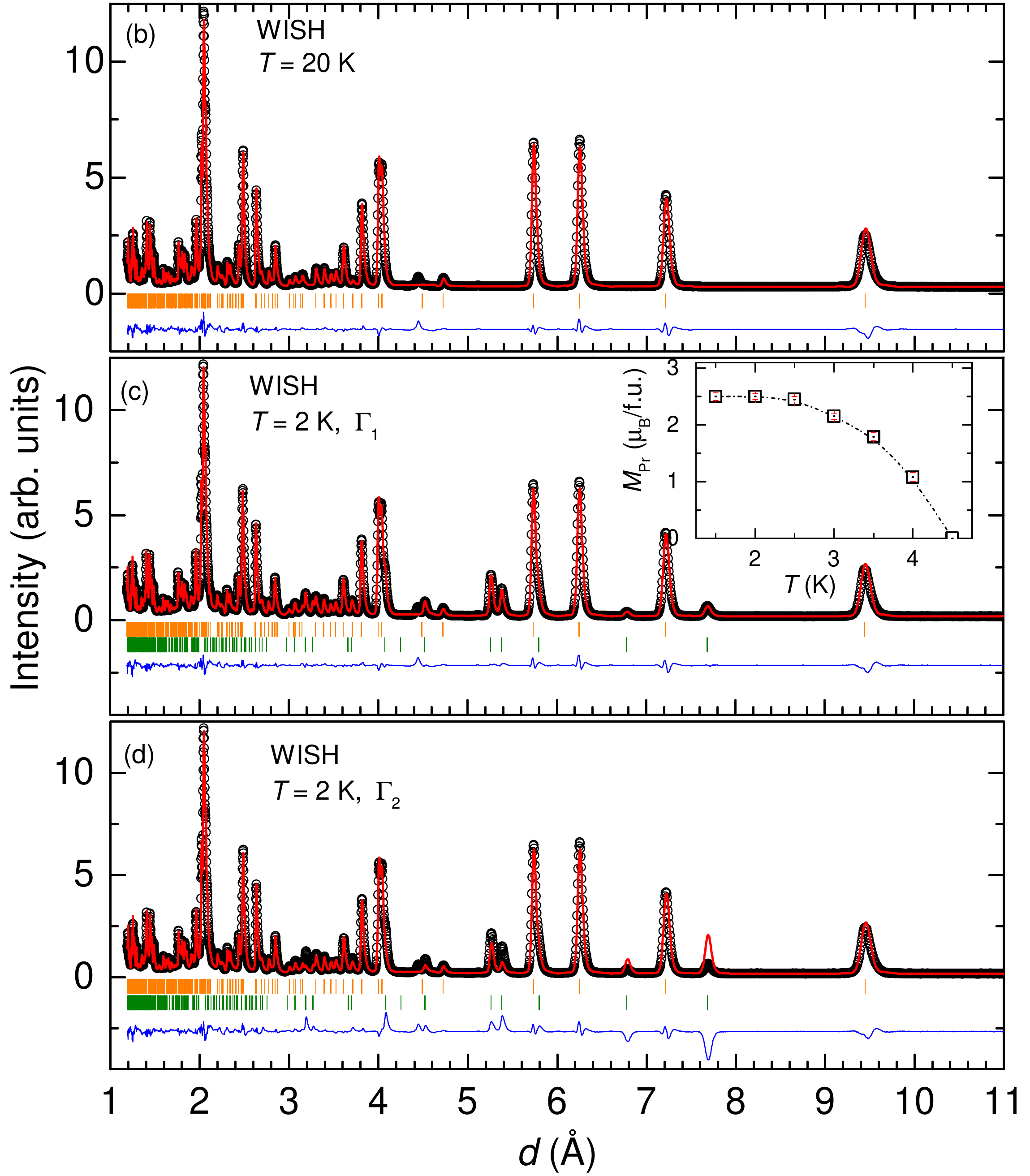}	
\caption{(color online) (a) Neutron powder diffraction intensity 
of \PFA\ at 25~K (circles) obtained from the time-of-flight 
diffractometer HRPD at ISIS facility, UK. Low temperature 
neutron diffraction pattern of \PFA\ at (b) 20~K and at (c) 
2~K obtained from the diffractometer WISH at ISIS facility, UK. 
A larger coverage of $d$-spacing helps in elucidating the magnetic 
structure of Pr in this compound. In panel (b), the lower set of 
vertical tick marks correspond to the Bragg peaks of the magnetic 
structure according to $\Gamma_1$ representation. The inset to (c) 
presents the temperature dependence of refined value of magnetic 
moment of Pr$^{3+}$ in \PFA\ showing a phase transition at 
$T^\mathrm{Pr}_N$ = 4.5~K. (d) The 2~K WISH data refined using 
the incorrect model $\Gamma_2$ does not account for the intensities 
correctly.\label{fig_wish}}
\end{figure}

A 2D contour plot of the diffracted intensity from 
\PFA\ obtained from the diffraction data of bank-3 of HRPD, 
is shown in figure~\ref{fig_hrpd_contour} as a 
function of temperature and $d$-spacing. 
Diffracted intensity due to Bragg scattering 
from the nuclear structure is present at 
$d \approx$ 4~{\AA}, 5.5~{\AA}, 6.2~{\AA}, 
7.2~{\AA} and 9.5~{\AA}. It is observed that 
additional magnetic intensity develops only 
below 4~K, at the $d$-spacing 5.3~{\AA}, 
which is indicated in figure~\ref{fig_hrpd_contour} 
by a horizontal  white dashed-line. Note that the 
temperature-axis is plotted in log-scale for clarity. 
According to our previous investigation on 
\PFA\ \cite{nair2016magnetic},~long-range magnetic 
order sets in only below $T^{Pr}_N\approx$ 4~K. 
The experimental evidence in figure~\ref{fig_hrpd_contour} 
supports this assignment of magnetic phase 
transition temperature $T^{Pr}_N\approx$ 4~K for Pr. 
In our previous work signatures akin to short-range 
spin fluctuations in the Fe lattice were observed in 
magnetic susceptibility at $T_{anom} \approx$ 34~K.  
No magnetic scattered intensity is developed at around 
34~K in the present diffraction data thereby ruling 
out the presence of magnetic ordering on the 
Fe lattice in this compound. More concretely, in a later subsection of the 
present paper, it is shown through inelastic neutron data that the energy-integrated
intensity as a function of $|Q|$ follows the magnetic form factor of
Pr$^{3+}$ there by confirming that only Pr orders magnetically in \PFA.
\\
%
\indent
The neutron powder diffraction intensity pattern of 
\PFA\ at 25~K obtained from HRPD, is plotted in 
figure~\ref{fig_wish} (a)  (black circles) as a function 
of $d$-spacing. The red solid line in the figure is the 
result of a Rietveld fit to the data using $Pbam$ 
space group. From the magnetization measurements
we know that at 25~K \PFA\ is not magnetically 
ordered and hence the diffraction data was modeled 
using only the nuclear structure in the space group 
model $Pbam$ as has been used in our previous work
\cite{nair2016magnetic}. The diffracted intensity at 300~K 
(not shown) is similar to that at 25~K as presented in 
figure~\ref{fig_wish} (a) and no signature of any 
structural phase transition is observed in this 
temperature interval of 300 - 25~K. The lattice 
parameters $a$, $b$ and $c$ show a gradual 
decrease with temperature as can be seen from 
the values at 300~K, 150~K and 25~K collected 
in Table~(\ref{tab1}). Also presented in 
Table~(\ref{tab1}) are the atomic parameters of \PFA\ at 300~K.
The crystal structure of \PFA\ in $Pbam$ space 
group has Pr occupying a single Wyckoff position 
$4g$\cite{tougait_jssc_2005prco}. 
The crystal structure is a close-packed arrangement 
of three dimensional Al polyhedra which are packed 
face-sharing to form chains of Pr
and Fe along the $c$-axis separated from each 
other by the Fe-Al network (see figure~\ref{fig_str}). 
The shortest Pr-Pr and Fe-Fe distances are parallel 
to the $c$-axis. As a function of temperature, the 
Pr-Pr and the Fe-Fe nearest-neighbour distances 
undergo only slight variation. A slight decrease in 
the nearest-neighbour distances is observed in all 
the directions $a$, $b$ and $c$. The bond angles 
between the nearest Pr atoms also do not vary 
significantly down till 1.6~K. Along $a$ and $b$ 
directions, they remain close to the values 
144$^\circ$ and 145$^\circ$ respectively. 
The bond parameters as a function of temperature 
are collected in Table~(\ref{tab2}).
\\
One of the important questions about the 
magnetism of \PFA\ is whether Pr and Fe in this 
compound order in a magnetically long-range 
fashion, either individually or jointly, 
or not. In our previous study\cite{nair2016magnetic}, 
magnetic anomalies at $T_{anom} \approx$ 34~K 
and $T^\mathrm{Pr}_N \approx$ 4~K were observed 
in the magnetization measurements. The anomaly at 
$T_{anom}$ was attributed to short-range order in 
the Fe lattice or due to minute amount of magnetic 
impurity; whereas $T^\mathrm{Pr}_N$ was attributed 
to long-range order in the Pr magnetic lattice.
Hence, one of the major aims of our neutron 
diffraction experiments was to determine the 
microscopic nature of the magnetic phase 
transitions pertaining to Fe or Pr in \PFA.
Since, through the low angle bank of HRPD, 
the development of long-range magnetic 
order in the Pr sublattice was observed to  occur 
below 4~K, we performed a detailed low temperature neutron 
diffraction measurements using the WISH 
diffractometer in the temperature range 2~K 
to 50~K covering a larger $d$-spacing up to 
11~{\AA}. The diffracted intensity at 50~K was identical to 
that of 20~K. No additional magnetic Bragg peaks 
or short range magnetic diffuse scattering 
were observed at 20~K.
Also, there was no enhancement of
diffracted intensity at low temperatures suggesting
ferromagnetic contributions at nuclear Bragg positions.
These results confirm that 
the 34~K transition observed previously\cite{nair2016magnetic} 
in the magnetic susceptibility is not from the bulk of the sample. 
\\
\begin{figure}[!t]
\centering
\includegraphics[scale=0.08]{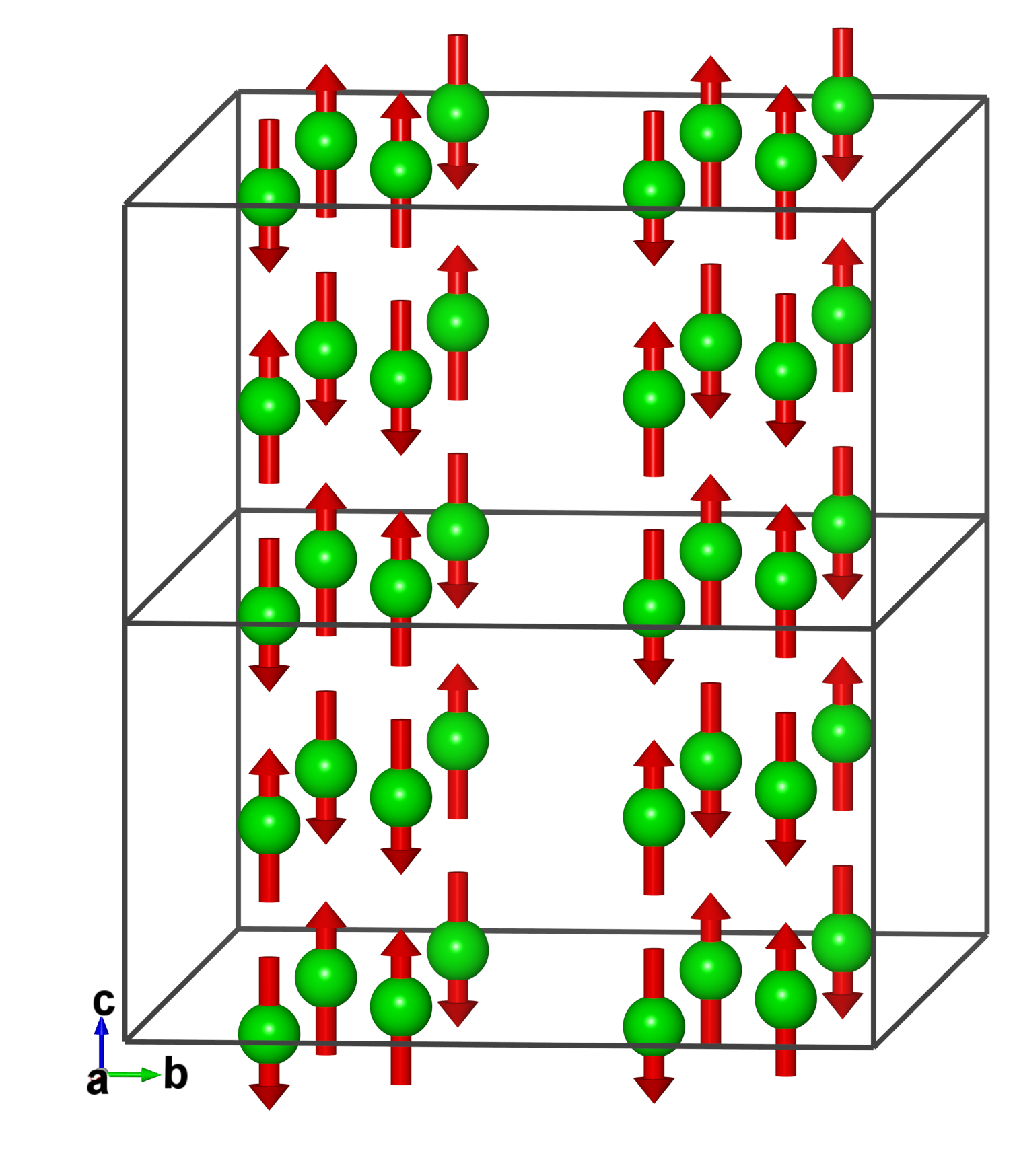}
\includegraphics[scale=0.08]{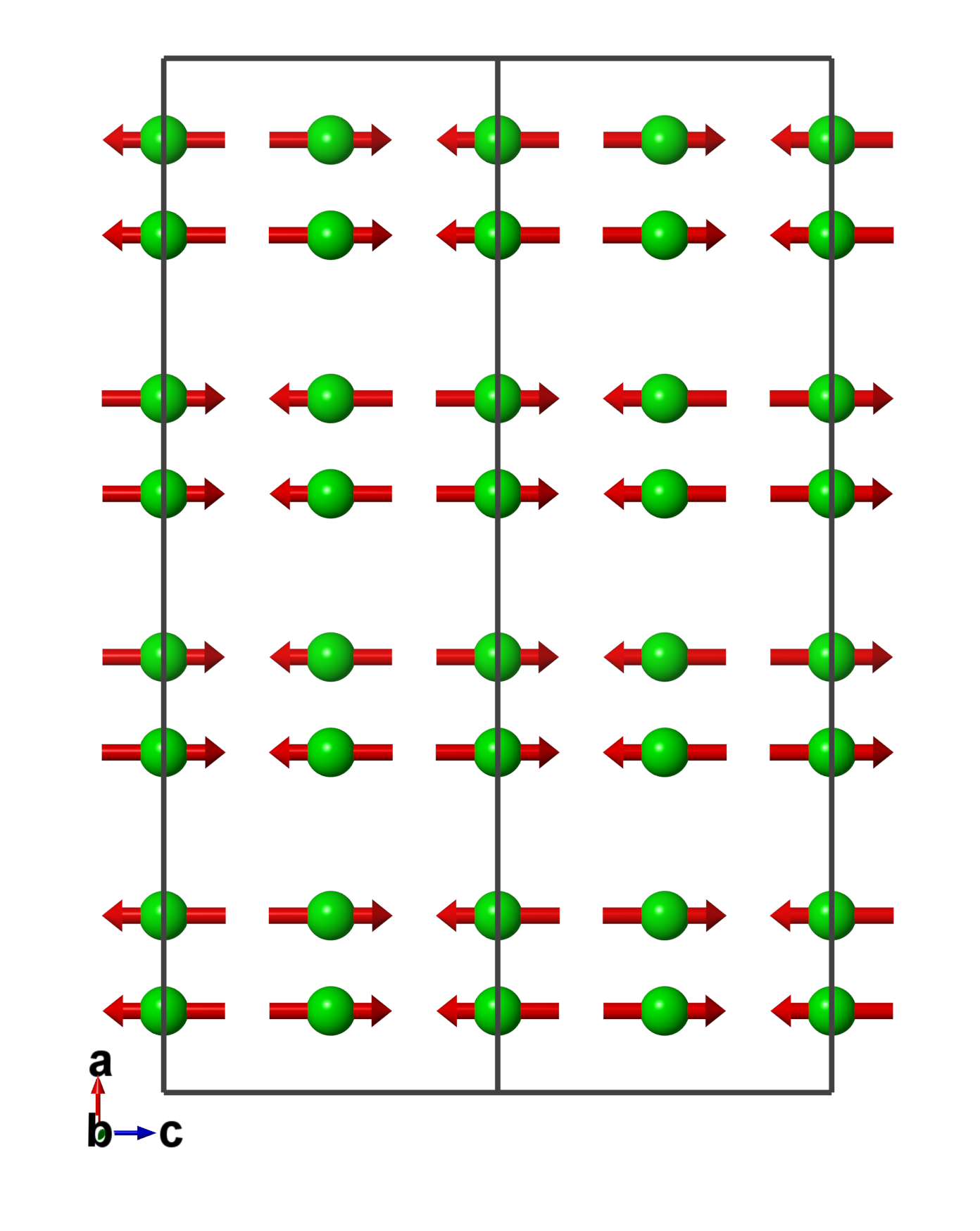}
\caption{(color online) The magnetic structure of \PFA\ 
according to the $\Gamma_1$ representation. The Pr$^{3+}$ 
moments are shown as red arrows ordering antiferromagnetically. 
Two different views of a 1$\times$1$\times$2 unit cell projected 
on to the plane are presented.\label{fig_Pr_Mag}}
\end{figure}
The neutron diffraction patterns at 20~K and 2~K 
obtained from WISH along with the Rietveld refinement results are 
presented in figure~\ref{fig_wish} (b), (c) and (d). 
It can be readily observed that 
additional scattered intensity has developed at 2~K 
as compared to the intensity profiles at high temperatures. 
Especially, new peaks at $d \approx$ 5.25~{\AA}, 
5.37~{\AA}, 6.78~{\AA} and 7.68~{\AA} emerge. 
The nuclear structure below $T^{Pr}_N \approx$ 
4~K conforms to $Pbam$, which is the crystal 
structure at 300~K. Hence, any structural phase 
transformation in the temperature range 300 - 4~K is ruled out.
In order to ascertain the ordering is from Pr or 
Fe or from both in \PFA\, first, a determination 
of the magnetic propagation vector was undertaken. 
Using the $k$-search utility in the FULLPROF suite of 
programs, the most probable propagation vector was 
determined as 
$\bf{k (\frac{1}{2}~0~\frac{1}{2})}$
to be used for the representation analysis of magnetic 
structure. According to the  symmetry analysis for the 
orthorhombic $Pbam$ space group, two magnetic 
representations $\Gamma_1$ and $\Gamma_2$ - 
both antiferromagnetic - were found as acceptable 
solutions for the magnetic structure of Pr$^{3+}$ in 
\PFA. Hence, Rietveld refinement of diffracted intensity 
at 2~K was performed by assuming the nuclear 
space group as $Pbam$ and the magnetic representation 
as $\Gamma_1$ or $\Gamma_2$. 
The magnetic $R$-factor for the refinements with $\Gamma_1$ or $\Gamma_2$ 
were 3.4 and 80 respectively, indicating that
$\Gamma_1$ is the best fit. In this representation, Pr 
moments order antiferromagnetically with moment 
direction along the orthorhombic $c$-axis. 
\begin{figure*}[!t]
	\centering
	\includegraphics[scale=0.30]{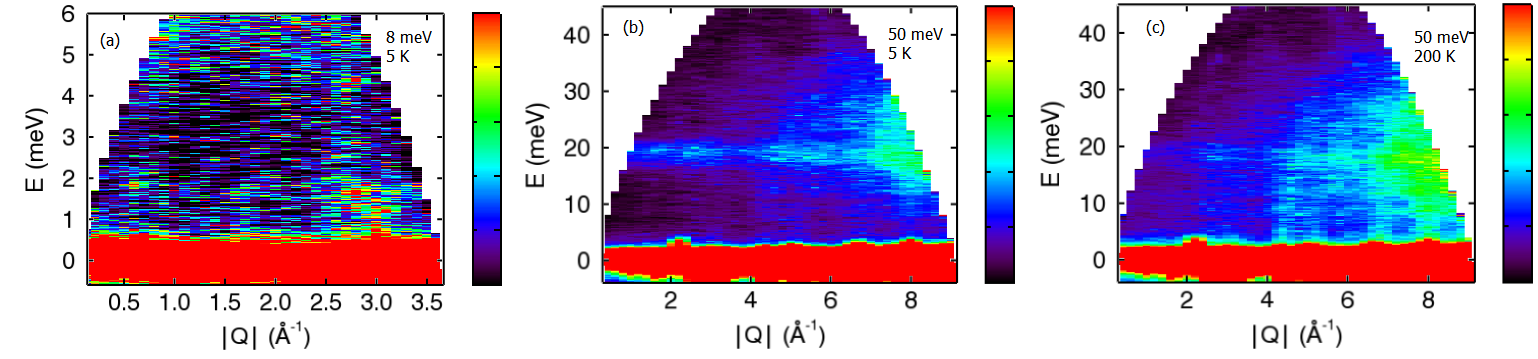}
	\includegraphics[scale=0.30]{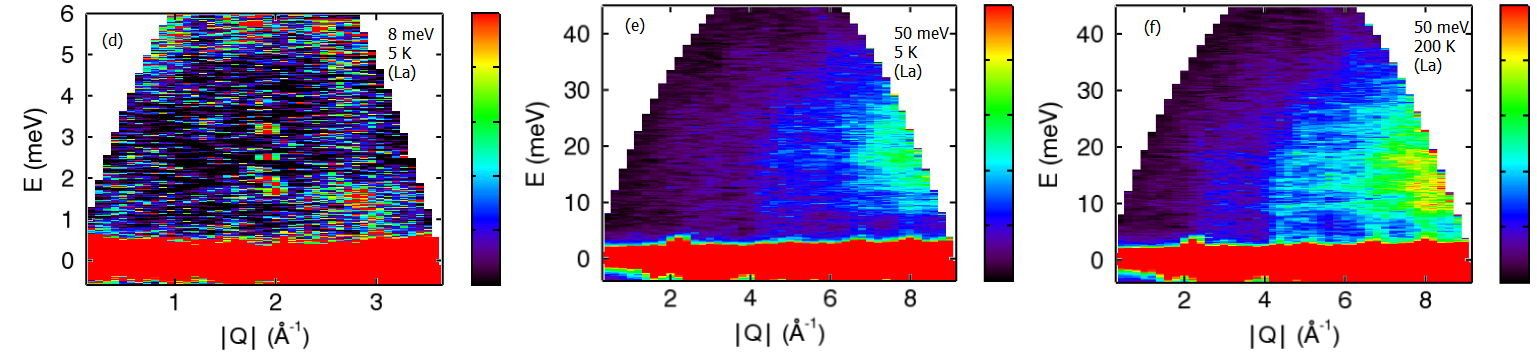}
	\caption{(color online) The color-coded plot of the inelastic neutron 
		scattering intensity of \PFA\ measured at MARI plotted as energy 
		transfer (E) versus momentum transfer ($\mathrm{|Q|}$) at (a) 5~K, 
		$E_i$ = 8~meV (b) 5~K, $E_i$ = 50~meV and (c) 200~K, $E_i$ = 50~meV; 
		all three for \PFA\, (d) 5~K, $E_i$ = 8~meV (e) 50~K, $E_i$ = 50~meV 
		and (f) 200~K, $E_i$ = 50~meV for LaFe$_2$Al$_8$. An inelastic 
		excitation is observed at $E \approx$ 19~meV for \PFA\ (in (b)) 
		which is attributed to crystal field effects. The intensity axes 
		were normalized to a scale of 50 in the units of mbr/sr/meV/fu.\label{fig_mari}}
\end{figure*}
In the panel (d) of Figure~\ref{fig_wish}, the Rietveld
refinement of 2~K data of \PFA\ is shown, with the fits
using $\Gamma_2$ representation (lower tick marks). Noticeably,
the intensities are not correctly accounted for in this model
(near $d \approx$ 5.5~\AA and 7.7~\AA).
A refinement model that allows for long-range magnetic order in 
both Pr and Fe lattices did not lead to an improved 
fit and hence was discarded. The experimental data 
did not fit to a model that considers only the Fe 
moment ordering, either. Moreover, from our previous 
experimental study\cite{nair2016magnetic} it was 
understood that the magnetic anomaly tentatively attributed
to Fe at $T_{anom}$ was very broad suggesting short-range correlations 
to be present rather than a conventional 
paramagnet-to-antiferromagnet phase transition.
A value of 2.5~$\mu_\mathrm{B}$/Pr is obtained at 2~K
for the neutron-derived ordered moment of Pr. A gradual decrease 
of the ordered magnetic moment was observed 
until $T^{Pr}_N$ = 4.5~K. The temperature-evolution 
of the ordered moment of Pr, $M_\mathrm{Pr}$, is 
presented in the inset of figure~\ref{fig_wish} (c). 
The dashed line in the inset serves as a guide to eye. 
The antiferromagnetic $\Gamma_1$ magnetic 
structure adopted by Pr$^{3+}$ moments in
\PFA\ is pictured in figure~\ref{fig_Pr_Mag}.
\subsection{Inelastic Neutron Scattering}
\begin{figure}[!t]
\centering
\includegraphics[scale=0.30]{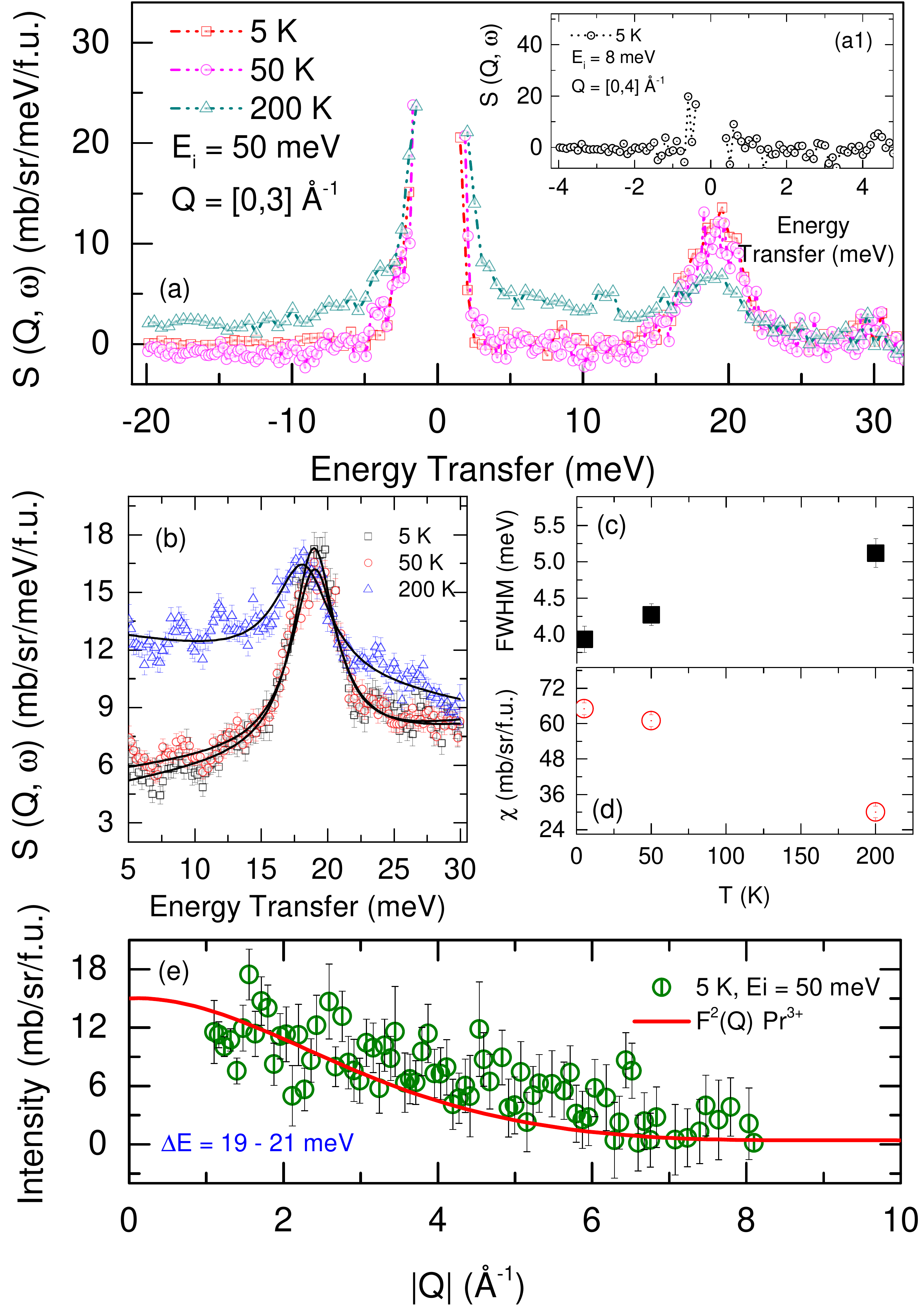}
\caption{(color online) (a) The dynamic structure factor $S(Q, \omega)$ 
of \PFA\ at 5~K, 50~K and 200~K plotted as a function of energy 
transfer ($E_i$ = 50~meV). The inset (a1) shows the $S(Q, \omega)$ 
at 5~K at an incident energy $E_i$ = 8~meV where no excitation is observed.
The curvefits (black lines), using a Bose-factor modified Lorentzian, 
to the 19~meV excitation at 5~K, 50~K and 200~K are shown in (b). 
The temperature dependence of FWHM and the local susceptibility, 
estimated from the profile fits in (b) are shown in (c) and (d) 
respectively. In (e)  the energy integrated intensity (19 to 21~meV) 
versus  $\mathrm{|Q|}$ is plotted along with the magnetic form factor, 
$F^2(Q)$, for Pr$^{3+}$. A peak observed near 12~meV in the 200~K data 
is possibly extrinsic in origin.\label{fig_Sqw}}
\end{figure}

Once we determined the magnetic structure of 
\PFA\, we now proceed to study the magnetic 
excitations and the crystal field effects using 
inelastic neutron scattering methods. The inelastic 
neutron scattering intensity of \PFA\ at different 
incident neutron energies and temperatures are 
presented in figure~\ref{fig_mari} (a) through (f) 
where color maps of energy transfer (in meV) 
versus momentum transfer ($|Q|$ in ${\AA^{-1}}$) 
are shown. Specifically, (a) $E_i$ = 8~meV, $T$ = 5~K, 
(b) $E_i$ = 50~meV, $T$ = 5~K and (c) $E_i$ = 50~meV, 
$T$ = 200~K for \PFA\ are shown. The inelastic 
intensity of the non-magnetic reference compound 
LaFe$_2$Al$_8$ are shown in (d) $E_i$ = 8~meV, 
$T$ = 5~K, (e) $E_i$ = 50~meV, $T$ = 5~K and (f) 
$E_i$ = 50~meV, $T$ = 200~K.
The intensity pattern at 5~K and at 200~K are 
essentially identical, suggesting that there is no
magnetic low lying excitations in the energy range
approximatley up to 10~meV.
The absence of any excited state transition at 200~K 
suggests that there are no CEF levels below 15~meV, 
which have non-zero matrix elements.
At low energy transfer, for example with $E_i$ = 8~meV 
(shown in (a)), no magnetic inelastic excitations are 
observed in \PFA. This result indicates that the 
quasi-elastic linewidth is very small and not able to 
detect within the resolution of MARI ($\Delta E$ = 
0.19~meV for $E_i$ = 8~meV, 150~Hz). In 
figure~\ref{fig_mari}, only one inelastic excitation is 
seen present at all the temperatures, {\em viz.,} at 
19~meV which arises 
from the crystal field excitation of the Pr$^{3+}$ 
crystal field levels, as can be seen from the panel (b) 
with incoming energy, $E_i$ = 50~meV
at 5~K. The inelastic spectra shown in figure~\ref{fig_mari} 
also suggest the absence of magnetic ordering 
down to 5~K, as one would expect spin wave 
excitations in an ordered state. Further the 
$|Q|$-dependent cuts (not shown) obtained by 
binning the energy from -2~meV to 2~meV (from (a) 
and (b)) did not reveal any magnetic Bragg peaks at 5~K. 
From panels (d, e) that represent the inelastic scattered 
intensity of LaFe$_2$Al$_8$ at 5~K, it is clear that no 
magnetic intensity is present in the non-magnetic analogue 
compound, which indicates that the Fe atoms do not have 
ordered magnetic moments in LaFe$_2$Al$_8$ or \PFA.
An $E$ versus $|Q|$ plot of \PFA\ obtained using 
$E_i$ = 150~meV (not shown here) also did not provide 
any evidence of other CEF excitations between 20~meV 
and 145~meV. The high intensity observed in (d) towards 
the high-$Q$ originates from the phonon contribution.
The inelastic intensity of \PFA\ at 200~K (in (c)) is 
compared with that of LaFe$_2$Al$_8$ in the panels (e) 
and (f), which further confirmed that high-$|Q|$ intensity 
is mainly dominated by the phonon scattering as its intensity 
shows an increase with increasing temperature.
\\
The $|Q|$-integrated (0 to 3~$\AA^{-1}$) dynamic 
structure factor, $S(Q, \omega)$, for \PFA\ as a 
function of energy transfer at $E_i$ = 50~meV is 
presented in figure~\ref{fig_Sqw} (a) for 5~K, 50~K 
and 200~K. The location in the graph where energy 
transfer equals zero corresponds to the
elastic line. The inelastic signal arising from the crystal 
field effect is observed at 19~meV which diminishes in 
intensity as temperature is raised. The $S (Q, \omega)$ 
dependence on energy transfer with $E_i$ = 8~meV is 
shown in the inset (a1), which shows the absence of any 
excitations in the energy range accessed. 
The dynamic structure factor for inelastic neutron
scattering of unpolarized neutrons is given by the
equation\cite{holland1982anomalous},
\begin{equation}
S (Q, \omega) = \frac{\hbar\omega}{1 - \mathrm{exp}(-\hbar\omega/k_\mathrm BT)}F^2(Q)\chi_0P(Q,\hbar\omega)
\label{Eqn:dynamicstrfactor}
\end{equation}
where $F(Q)$ is the magnetic form factor of the magnetic ion,
$\chi_0$ is the static magnetic susceptibility and $P(Q, \hbar\omega)$ is
a spectral function. The first term on the right-hand-side
is the Bose factor for the thermal occupation of levels.
Curve fits to the 19~meV excitation according to
Equation~(\ref{Eqn:dynamicstrfactor}) is shown in
Figure~\ref{fig_Sqw} (b) for 5~K, 50~K and 200~K.
The FWHM of the peaks as well as the area under the peak
which corresponds to the local susceptibility were
extracted from these fits and presented respectively 
in panels (c) and (d) of Figure~\ref{fig_Sqw}.
A gradual broadening of the peaks with increasing temperature
is observed, suggested by the increasing FWHM and decreasing
susceptibility. A peak is observed at $E \approx$ 12~meV for the 
200~K data (in the main panel of (a)) which, in fact,
is narrower than the resolution and hence is judged
as a spurious peak.
The $|Q|$-dependence of integrated intensity at 5~K, 
binned in the interval of 19 - 21~meV is shown in the 
panel (e) plotted together with
the square of the magnetic form factor, $F^2(Q)$ for 
Pr$^{3+}$. The integrated 
intensity follows the $F^2(Q)$ for Pr$^{3+}$ closely,
suggesting that the inelastic excitation in \PFA\
results from single ion effects of Pr$^{3+}$. In 
order to estimate the magnetic moment from the  
inelastic signal at 19~meV, use was made of the second sum rule:
\begin{equation}
\int_{-\omega}^{\omega} S(Q,\omega)/F^2(Q) d\omega = A\mu^2_\mathrm{eff}
\end{equation}
where $S (Q, \omega)$ is the dynamic 
structure factor, $F (Q)$ is the magnetic 
form factor, $A$ = 48.6 mb/sr/$\mu^2_\mathrm{B}$
is a constant\cite{holland1982anomalous} 
and $\mu_\mathrm{eff}$ 
is the effective magnetic moment. From the 
integrated intensity at the 19~meV peak of \PFA\,
the effective moment value was estimated as 
$\mu_\mathrm{eff}$ = 1.6(2)~$\mu_\mathrm{B}$ 
which is almost half of the spin-orbit coupled 
magnetic moment $\mu^\mathrm{Pr}_\mathrm{so}$ 
= 3.58~$\mu_\mathrm{B}$ of Pr$^{3+}$, $J$ = 4. 
The low value of the moment associated with 19~meV 
excitation along with the heat capacity results, which 
suggest a doublet CEF ground state, indicates that the 
missing Pr moment is associated with a narrow resolution 
limited quasi-elastic scattering from the ground state. 
%
%
\section{Discussion}
The specific heat of \PFA~ showed a clear indication 
of a peak at $T^\mathrm{Pr}_N \approx$ 4~K where 
the Pr begins to order magnetically\cite{nair2016magnetic}. 
The magnetic contribution to the specific heat was 
obtained by subtracting the non-magnetic specific 
heat of LaFe$_2$Al$_8$. The estimate of crystal 
field levels were obtained by modeling the Schottky 
peak using a 9-level model for the crystal field levels. 
The first excited crystal field level was calculated at 
$\approx$ 10~K and from the behaviour of magnetic 
entropy as a function of temperature, a full occupation 
of the 9 levels was found to happen at 
$\approx$ 200~K where the entropy reaches $R$~ln(9).
From the inelastic neutron scattering data, 
a single excitation is discernible at 19~meV 
which corresponds to $\approx$ 200~K.
The first excited level near 10~K ($\approx$ 0.8~meV) 
is not observed in the inelastic data  with either 
$E_i$ = 8~meV or 50~meV, which could be due to 
the zero matrix  element between these wave functions - 
the intensity of CEF excitation vary as a square of this matrix element.
However, since only one excitation is revealed in 
the inelastic scattering experiment, a full crystal field 
analysis to estimate all the 9 crystal field levels is not 
possible in the present case. Considering orthorhombic 
point symmetry ($C_s$) of Pr ion in \PFA\, there will be 
nine non-zero crystal field parameters in the Hamiltonian 
and hence a single observed CEF excitation is not 
sufficient to fit the CEF model.
The temperature dependence of the inelastic 
excitation peak at 19~meV shown in figure~\ref{fig_Sqw} (a) 
indicates that the peak position does not shift with temperature, 
confirming the origin of the excitation to be crystal field effect. 
Moreover, the integrated intensity at 19~meV as a function of 
$Q$ follows the square of form factor $F^2(Q)$ of 
Pr$^{3+}$ confirming that the magnetism in
\PFA\ is purely from Pr$^{3+}$. Though the previous 
magnetization study of \PFA\ indicated short-range 
magnetic order due to Fe, the neutron scattering 
results (figure~\ref{fig_hrpd_contour}) do not show 
any indication of this. The diffraction results only 
indicate that the long-range magnetic ordering in \PFA\ 
at 4.5~K pertains only to the Pr sublattice where the moments order 
antiferromagnetically.
The crystal field level scheme for $J$ = 4, 
Pr$^{3+}$ in the local symmetry of $m$ predicts 
a singlet ground state. The 2$J$ + 1 = 9 fold multiplet 
of Pr$^{3+}$ splits so that a crystal field singlet is 
always present for all type of crystallographic 
symmetry\cite{walter_treating_1984}.  Pr-based 
intermetallics often display singlet ground states 
however, also
have been found to lead to exotic behaviour 
such as the observation of heavy fermion 
superconductivity\cite{bauer_superconductivity_2002,goremychkin_crystal_2004}. 
Not only that, in certain Pr based systems like 
PrAu$_2$Si$_2$-related compounds, frustration 
effects on the magnetic ground state were argued 
to be introduced by crystal field effects\cite{goremychkin2008spin}. 
In fact, a competition between the crystal field interaction 
and the magnetic exchange energy strengths were 
observed in these compounds, eventually, 
resulting in a novel spin glass ground state. 
Magnetic long range order also happens in several 
systems despite the crystal field level scheme 
warranting otherwise. PrIrSi$_3$ is one such 
system where Pr displays long range magnetic 
order\cite{anand2014investigations}. However a 
large separation of about 90~K is observed between 
the first excited crystal field levels of PrIrSi$_3$. 
Induced moment magnetic ordering in this compound 
would require very high internal exchange energy to 
achieve ordering. In some other Pr-based compounds 
like PrRu$_2$Si$_2$ and PrNi$_2$Si$_2$, 
relatively smaller crystal field separations are 
(of 2.25~meV and 3.3~meV, respectively)
\cite{mulders_prru2si2_1997,blanco_specific_1992}. 
In the present case of \PFA\, the knowledge 
of the first excited crystal field level first originated from a simulation 
of the Schottky specific heat, which predicts the first level at around 10~K. 
An estimation of the exchange constant is in general 
possible through the relation, 
$\theta_p$ = $\left[\frac{J_\mathrm{ex}J(J + 1)}{3k_\mathrm B}\right]$ 
where $\theta_p$ is the Curie-Weiss temperature 
and $J_\mathrm{ex}$ is the exchange constant. 
However, in the case of \PFA\ it was found that the 
inverse magnetic susceptibility did not follow a 
linear trend due to the presence of Fe short-range 
fluctuations and hence a mean-field estimate of 
$\theta_p$ was not possible. Comparing the ground state
energy scale in \PFA\ with other 
compounds like PrIrSi$_3$ or with PrRu$_2$Si$_2$ 
and PrNi$_2$Si$_2$, it seems more plausible that an 
exchange-generated admixture of low lying crystal field 
energy levels is possible in the case of \PFA.
\section{Conclusions}
In conclusion, through neutron powder diffraction 
and inelastic scattering experiments we determined 
the magnetism of the quasi-skutterudite compound 
\PFA\. Through the analysis of the neutron diffraction 
data, it is found that only Pr develops long-range 
magnetic order in this compound with a 
$T^\mathrm{Pr}_N$ = 4.5~K. The Pr$^{3+}$ 
moments adopt antiferromagnetic arrangement 
with moments along the $c$-axis and a propagation 
vector $\bf{k (\frac{1}{2}~0~\frac{1}{2})}$. Only one 
magnetic excitation at 19~meV is observed in the 
inelastic neutron scattering data of \PFA\, 
corresponding to the crystal field excitation from 
Pr$^{3+}$ levels. The presence of only one 
excitation detected in our experiments has rendered 
the estimation of the full crystal field level scheme of 
Pr$^{3+}$ in \PFA\ impossible  as there are nine 
free crystal field parameters to be determined from 
the one excitation. However, from the $|Q|$-dependence 
of integrated magnetic intensity it is confirmed that 
the magnetism in \PFA\ is solely governed by Pr 
while the Fe sublattice remains magnetically 
inactive down to 2~K.
\section*{Acknowledgements}
HSN acknowledges FRC/URC for a postdoctoral fellowship. 
AMS thanks the SA-NRF (93549) and the FRC/URC of UJ for financial assistance. 
DTA thanks  CMPC-STFC grant number CMPC-09108 for the funding.
We would like to thank A. Bhattacharyya for his help on MARI experiment.

\begin{thebibliography}{10}
	\expandafter\ifx\csname url\endcsname\relax
	\def\url#1{{\tt #1}}\fi
	\expandafter\ifx\csname urlprefix\endcsname\relax\def\urlprefix{URL }\fi
	\providecommand{\eprint}[2][]{\url{#2}}
	
	\bibitem{sales_filled_1996}
	Sales B~C, Mandrus D and Williams R~K 1996 {\em Science\/} {\bf 272} 1325
	
	\bibitem{nolas_skutterudites:_1999}
	Nolas G~S, Morelli D~T and Tritt T~M 1999 {\em Ann. Rev. Mater. Sci.\/} {\bf
		29} 89--116
	
	\bibitem{guo_development_2012}
	Guo J~Q, Geng H~Y, Ochi T, Suzuki S, Kikuchi M, Yamaguchi Y and Ito S 2012 {\em
		J. Elec. Mater.\/} {\bf 41} 1036--1042
	
	\bibitem{frank_complex_1958}
	Frank F~C and Kasper J~S 1958 {\em Acta Crystallogr.\/} {\bf 11} 184--190
	
	\bibitem{tursina_ceru2al10_2005}
	Tursina A~I, Nesterenko S~N, Murashova E~V, Chernyshev I~V, Noël H and
	Seropegin Y~D 2005 {\em Acta Crystallogr. E\/} {\bf 61} i12--i14
	
	\bibitem{adroja_muon-spin-relaxation_2013}
	Adroja D~T, Hillier A~D, Muro Y, Takabatake T, Strydom A~M, Bhattacharyya A,
	Daoud-Aladin A and Taylor J~W 2013 {\em Physica Scripta\/} {\bf 88} 068505
	
	\bibitem{takeda_superconducting_2000}
	Takeda N and Ishikawa M 2000 {\em J. Phys. Soc. Jpn.\/} {\bf 69} 868--873
	
	\bibitem{adroja_probing_2005}
	Adroja D~T, Hillier A~D, Park J~G, Goremychkin E~A, McEwen K~A, Takeda N,
	Osborn R, Rainford B~D and Ibberson R~M 2005 {\em Physical Review B\/} {\bf
		72} 184503
	
	\bibitem{ito_sr_2011}
	Ito T~U, Wataru H, Kazuhiko N, Hubertus L, Christopher B, Akito S and Satoru N
	2011 {\em J. Phys. Soc. Jpn.\/} {\bf 80} 113703
	
	\bibitem{sakoda_single_2012}
	Masahito S, Kazuhiro K, Shuhei T, Eiichi M, Hitoshi S, Matsuda T~D and
	Yoshinori H 2012 {\em J. Phys. Soc. Jpn.\/} {\bf 81} SB011
	
	\bibitem{slebarski2015study}
	{\'S}lebarski A, Goraus J, Witas P, Kalinowski L and Fija{\l}kowski M 2015 {\em
		Phys. Rev. B\/} {\bf 91} 035101
	
	\bibitem{khuntia2014contiguous}
	Khuntia P, Peratheepan P, Strydom A~M, Utsumi Y, Ko K~T, Tsuei K~D, Tjeng L~H,
	Steglich F and Baenitz M 2014 {\em Phys. Rev. Lett.\/} {\bf 113} 216403
	
	\bibitem{snyman2012anomalous}
	Snyman J~L and Strydom A~M 2012 {\em J. Appl. Phys.\/} {\bf 111} 07A943
	
	\bibitem{nair2016magnetic}
	Nair H~S, Ghosh S~K, Rameshkumar K and Strydom A~M 2016 {\em J. Phys. Chem.
		Solids\/} {\bf 91} 69--75
	
	\bibitem{dave}
	Azuah R~T, Kneller L~R, Qiu Y, Tregenna-Piggott P~L~W, Brown C~M, Copley J~R~D
	and Dimeo R~M 2009 {\em J. Res. Nat. Inst. Standards and Tech.\/} {\bf 114}
	341
	
	\bibitem{rietveld}
	Rietveld H~M 1969 {\em J. Appl. Crystall.\/} {\bf 2} 65--71
	
	\bibitem{fullprof}
	Rodriguez-Carvajal J 2010 {\em LLB, CEA-CNRS, France [http://www. ill.
		eu/sites/fullprof/]\/}
	
	\bibitem{sarah}
	Wills A~S 2000 {\em Physica B\/} {\bf 276} 680--681
	
	\bibitem{tougait_jssc_2005prco}
	Tougait O, Kaczorowski D and No{\"e}l H 2005 {\em J. Solid State Chem.\/} {\bf
		178} 3639--3647
	
	\bibitem{holland1982anomalous}
	Holland-Moritz E, Wohlleben D and Loewenhaupt M 1982 {\em Phys. Rev. B\/} {\bf
		25} 7482
	
	\bibitem{walter_treating_1984}
	Walter U 1984 {\em J. Phys. Chem. Solids\/} {\bf 45} 401--408
	
	\bibitem{bauer_superconductivity_2002}
	Bauer E~D, Frederick N~A, Ho P~C, Zapf V~S and Maple M~B 2002 {\em Phys. Rev.
		B\/} {\bf 65} 100506
	
	\bibitem{goremychkin_crystal_2004}
	{E A Goremychkin}, Osborn R, Bauer E~D, Maple M~B, Frederick N~A, Yuhasz W~M,
	Woodward F~M and Lynn J~W 2004 {\em Phys. Rev. Lett.\/} {\bf 93} 157003
	
	\bibitem{goremychkin2008spin}
	Goremychkin E~A, Osborn R, Rainford B~D, Macaluso R~T, Adroja D~T and Koza M
	2008 {\em Nat. Phys.\/} {\bf 4} 766--770
	
	\bibitem{anand2014investigations}
	Anand V~K, Adroja D~T, Bhattacharyya A, Hillier A~D, Taylor J~W and Strydom A~M
	2014 {\em J. Phys.: Condens. Matter\/} {\bf 26} 306001
	
	\bibitem{mulders_prru2si2_1997}
	{A M Mulders}, Yaouanc A, Dalmas~de Réotier P, Gubbens P~C~M, Moolenaar A~A,
	Fåk B, Ressouche E, Prokeš K, Menovsky A~A and Buschow K~H~J 1997 {\em
		Phys. Rev. B\/} {\bf 56} 8752--8759
	
	\bibitem{blanco_specific_1992}
	Blanco J~A, Gignoux D and Schmitt D 1992 {\em Phys. Rev. B\/} {\bf 45}
	2529--2532
	
\end{thebibliography}

\providecommand{\newblock}{}

\end{document}